\documentclass[conference,a4paper]{IEEEtran}
\IEEEoverridecommandlockouts
\usepackage{soul} 
\usepackage{cite}
\usepackage{amsmath,amssymb,amsfonts}
\usepackage{algorithmic}
\usepackage{graphicx}
\usepackage{textcomp}
\usepackage{xcolor}
\usepackage{lipsum}
\def\BibTeX{{\rm B\kern-.05em{\sc i\kern-.025em b}\kern-.08em T\kern-.1667em\lower.7ex\hbox{E}\kern-.125emX}}

\usepackage{booktabs}
\usepackage{makecell}

\usepackage{balance}

\title{Tunable Ferroelectric Acoustic Resonators in \\Monolithic Thin-Film Barium Titanate}

\author{\IEEEauthorblockN{Ian Anderson\IEEEauthorrefmark{1},
Agham Posadas\IEEEauthorrefmark{3},
Alexander A. Demkov\IEEEauthorrefmark{2}, and
Ruochen Lu\IEEEauthorrefmark{1}}
\IEEEauthorblockA{\IEEEauthorrefmark{1}Department of Electrical and Computer Engineering, The University of Texas at Austin, Austin, TX, USA }
\IEEEauthorblockA{\IEEEauthorrefmark{2}Department of Physics, The University of Texas at Austin, Austin, TX, USA}
\IEEEauthorblockA{\IEEEauthorrefmark{3}La Luce Cristallina, Inc., Austin, TX, USA}
\IEEEauthorblockA{ianderson@utexas.edu}}

\begin{document}
\bstctlcite{IEEEexample:BSTcontrol}

\maketitle

\begin{abstract}
The increasing development of wireless communication bands has motivated the development of compact, low-loss, and frequency adjustable RF filtering technologies. Acoustic resonators are the ideal solution to these requirements, and tunable implementations offer a path toward reconfigurable front ends. In this work, we investigate epitaxial barium titanate (BTO) grown on silicon as a platform for tunable acoustic resonators operating in the sub-GHz regime. We demonstrate lateral excitation of symmetric lamb (S0) modes in X-cut BTO membranes, in contrast to prior thickness-defined ferroelectric resonators. Devices are designed using finite-element simulations and fabricated with laterally patterned electrodes that enable overtone coupling to multiple resonant modes. Under applied DC bias, ferroelectric domains align, allowing electrical excitation, frequency tuning, and quality-factor enhancement of acoustic modes. Resonances near 300 MHz and 700 MHz exhibit electromechanical coupling up to 8\% and bias-dependent frequency tuning, with a distinct transition in behavior near 20 V. These results highlight monolithic BTO on silicon as a promising material system for laterally excited, tunable acoustic resonators for reconfigurable RF applications.
\end{abstract}


\section{Introduction}
Modern wireless communications and technology have progressively shifted toward smaller, more clustered frequency bands with higher data rates\cite{attar_5g_2022}. With each additional communication band comes the addition of more Radio Frequency (RF) components to accommodate these bands and selectively select one signal from hundreds. Acoustic filters, with small size and low insertion loss, are ideal candidates for these tasks, as one can fit far more acoustic filters into cellular devices than electromagnetic (EM) versions \cite{hagelauer_microwave_2023,jackson_optical_1985}.

Current acoustic technology assigns one filter per frequency band to select the appropriate information. Thin film piezoelectric materials commonly used include aluminum nitride (AlN), scandium aluminum nitride (ScAlN), Lithium Niobate (LN), and Lithium Tantalate (LT) \cite{qian_heterogeneous_2025,rinaldi_super-high-frequency_2010,lu_rf_2021}. However, an alternative to using one filter per band is to use a single tunable filter across multiple frequency bands. Technologies for tunable integrated resonators/filters include phase change materials\cite{hummel_reconfigurable_2015}, ferromagnetics\cite{devitt_spin-wave_2026}, or MEMS varactors and switches\cite{konno_tunable_2013, nordquist_off_2013}. Ferroelectrics offer an alternative route to integration, requiring only a DC bias in addition to the AC signal for tuning. These materials use a tuning of electromechanical coupling ($k^2$), or change in effective stiffness to change resonance frequency, and thus change filter frequency or turn off the filter altogether\cite{alzuaga_tunable_2014}. The most commonly used ferroelectric material in the acoustic domain is ScAlN, but is limited by only changing frequency with changes in effective stiffness, and the device cannot be turned on and off \cite{wang_film_2020,fichtner_alscn_2019}. Barium Titanate (BTO), alongside its lower Curie temperature counterpart Barium Strontium Titanate (BST), is an excellent candidate for tunable filters. Compared to ScAlN, the tunability is far greater through the increase of coupling with DC bias. BTO also offers several other advantages, including epitaxial growth on silicon (Si), control of orientation via growth conditions, and integration with other device types, such as electro-optic modulators \cite{kim_crystal_2025,dong_monolithic_2023,kim_nature_2023}. However, prior BTO/BST resonators typically rely on thickness-defined frequencies and often require bottom electrodes, which limits monolithic integration and makes multi-frequency-on-chip difficult without changing film thickness or adding additional process complexity.

\begin{figure}
\centering
\includegraphics[width=3.2in]{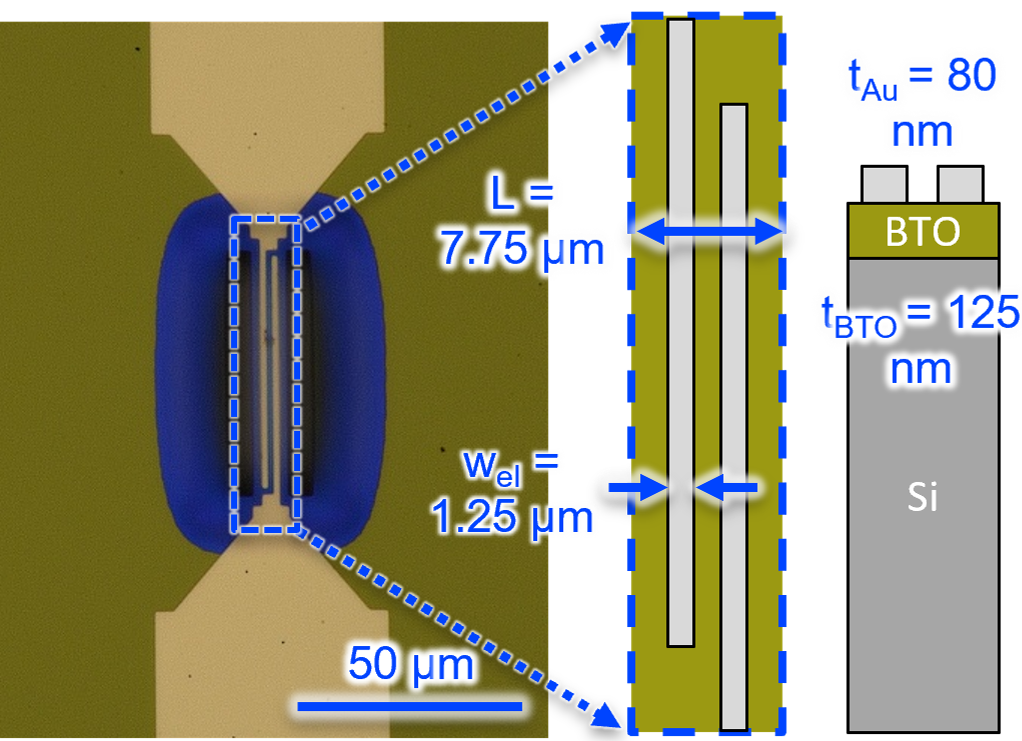}
\caption{Device optical image showing dimensions and electrode layout. }
\label{fig1}
\end{figure}

The following work focuses on Epitaxial BTO on Si as a material for tunable acoustic resonators. Previous demonstrations of BTO have focused on film bulk acoustic resonators (FBARs) for acoustic devices, utilizing thickness electrical-field profiles to excite acoustic modes\cite{koohi_high_2017,lee_intrinsically_2013,kongbrailatpam_switchable_2025}. Here, we demonstrate lateral excitation of symmetric Lamb modes in X-Cut BTO in the sub-GHz range.

\section{Design and Simulation}

Simulations were performed in COMSOL Multiphysics to determine optimal electrode configurations for coupling to the following mode profiles. Simulations are performed with 125 nm of BTO and 75 nm of gold for electrodes. Due to high film stress, release conditions were limited to isotropic etching of approximately 10 $\mu$m of silicon laterally to limit the out-of-plane deflection of said devices, as can be seen by the gradient of color in the blue released region. A total lateral size of 7.75 $\mu$m was chosen, with an electrode size of 1.25 $\mu$m and an aperture of 50 $\mu$m. Devices utilize the $e_{11}$ coefficient to excite fundamental symmetric lamb modes (S0). Due to the lateral spacing between the electrodes and the etch windows, the device functions as an overtone resonator and couples to multiple modes rather than a single mode \cite{lee_nanoscale_2023}. The admittance plot of the device is shown in Fig. 2, which depicts the different modes we are coupling into, with progressively higher-order stress nodes.

\begin{figure}
\centering
\includegraphics[width=\columnwidth]{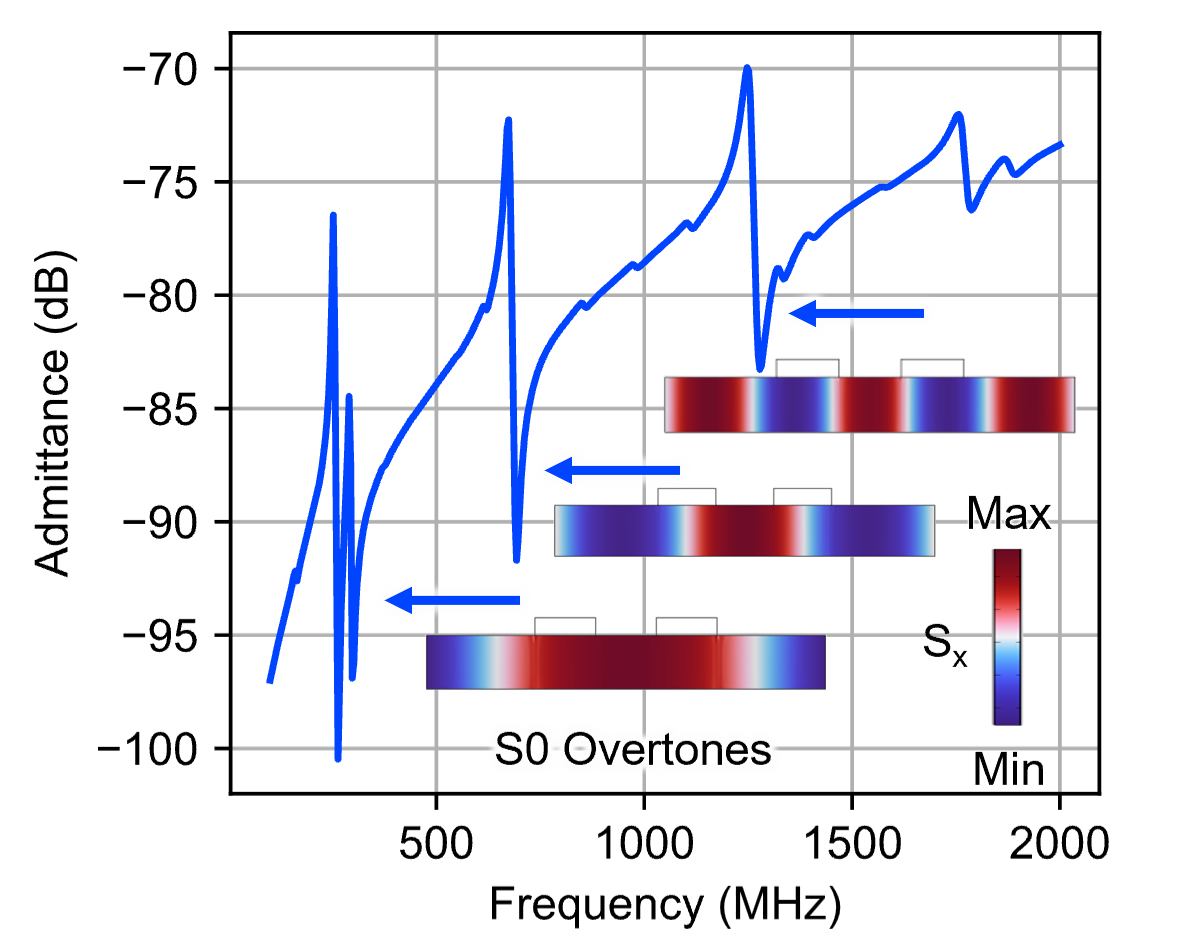}
\caption{COMSOL admittance simulation shows S0 overtones and their stress profiles. }
\label{fig1}
\end{figure}

\section{Measurement and Analysis}

\begin{figure}
\centering
\includegraphics[width=\columnwidth]{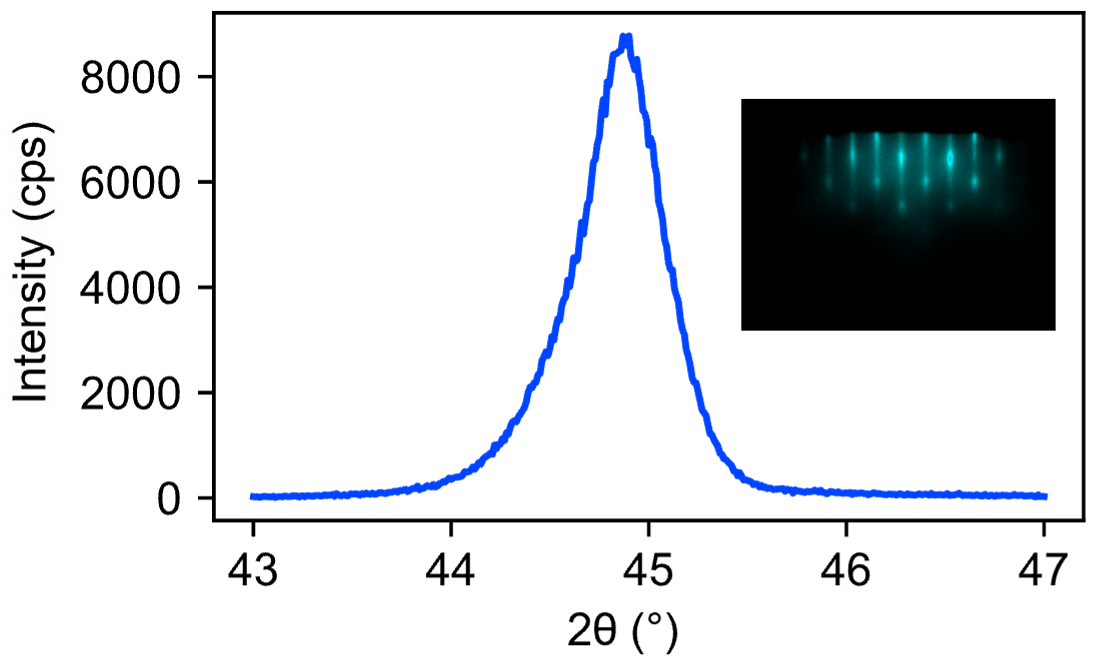}
\caption{XRD 2$\theta$ scan measurements showing peak slightly lower than 45 degrees, with inset of RHEED pattern.}
\label{fig_xrd}
\end{figure}

For this study, intrinsic silicon wafers (R$\approx$ 10000 $\Omega$-cm) with 2” diameter were used as substrates. Before BTO deposition, a 5 nm-thick SrTiO$_3$ (STO) buffer layer was deposited on the clean Si surface by molecular beam epitaxy and subsequently transferred under vacuum to a sputtering system. The BTO layer was deposited by off-axis RF magnetron sputtering at a substrate temperature of 700°C and was grown to a thickness of 120 nm. Epitaxial growth was confirmed using reflection high-energy electron diffraction (RHEED) and X-ray diffraction (XRD) in Fig. 3. The 120-nm BTO films on bulk Si showed an out-of-place lattice constant of 4.036 \r{A}.

\begin{figure}
\centering
\includegraphics[width=\columnwidth]{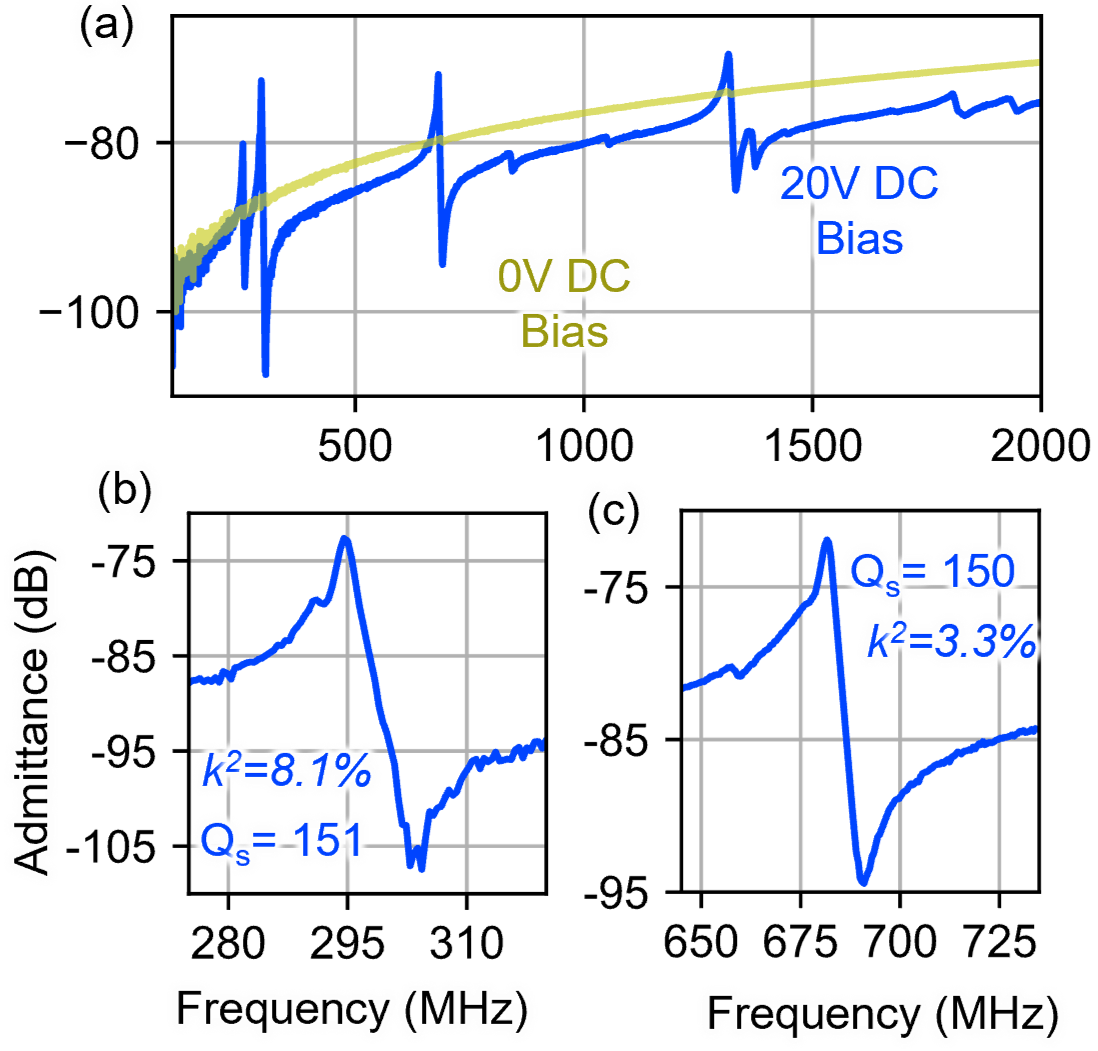}
\caption{(a) 0V versus 20V DC Bias wide admittance measurement (b) zoom of mode at 300 MHz showing $k^2$ and $Q$, and (c) same zoom of mode at 700 MHz.}
\label{fig1}
\end{figure}

Devices were measured using a Vector Network Analyzer (VNA) with an applied DC bias between ports 1 and 2 \cite{nordlander_ferroelectric_2020}. When the stack is grown, unit cells have spontaneous polarizations that can point in one of four directions for a-axis BTO, termed ferroelectric domains \cite{vasudevan_domain_2023}. For this reason, generally, electromechanical coupling cancels out, and no modes can be seen for unbiased measurements. However, when a DC bias is overlaid with our AC signal, these unit cells align, a net piezoelectric coefficient is realized, and acoustic modes can be excited with nonzero coupling. Fig. 4 shows examples of our device measured under 0 V and 20 V DC bias, indicating that our modes exist only under external bias. Here, we present our previously simulated modes with electromechanical coupling of 8$\%$ and 3$\%$, and quality factor (Q) of 150 for each. The different modes arise from the gap between electrodes and release windows, causing overtone coupling.

\begin{figure}
\centering
\includegraphics[width=\columnwidth]{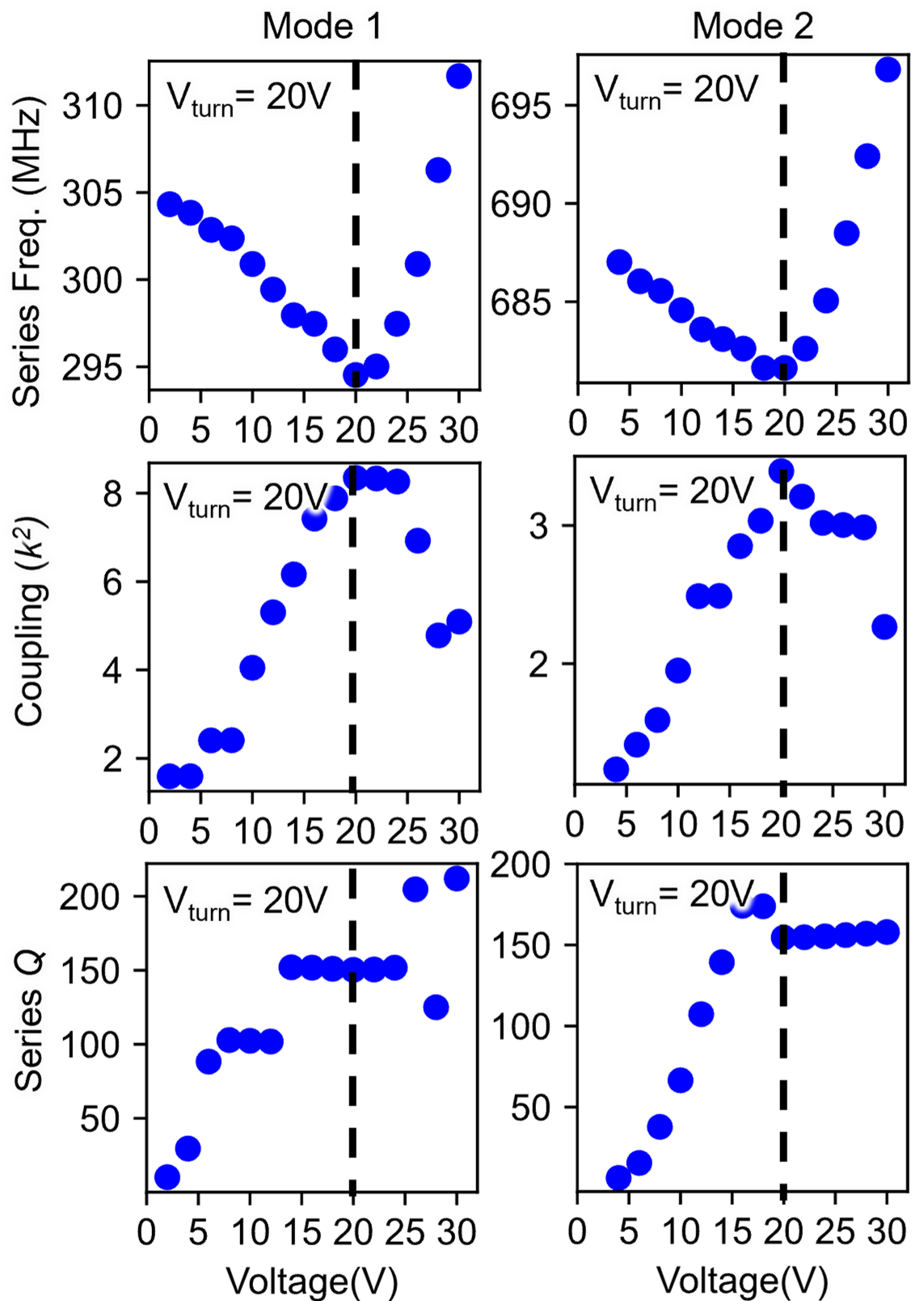}
\caption{Mode 1 at 300 MHz and mode 2 at 700 MHz behaviors versus applied DC bias showing electromechanical coupling, series quality factor, and series resonance frequency.}
\label{fig1}
\end{figure}

\begin{figure}
\centering
\includegraphics[width=\columnwidth]{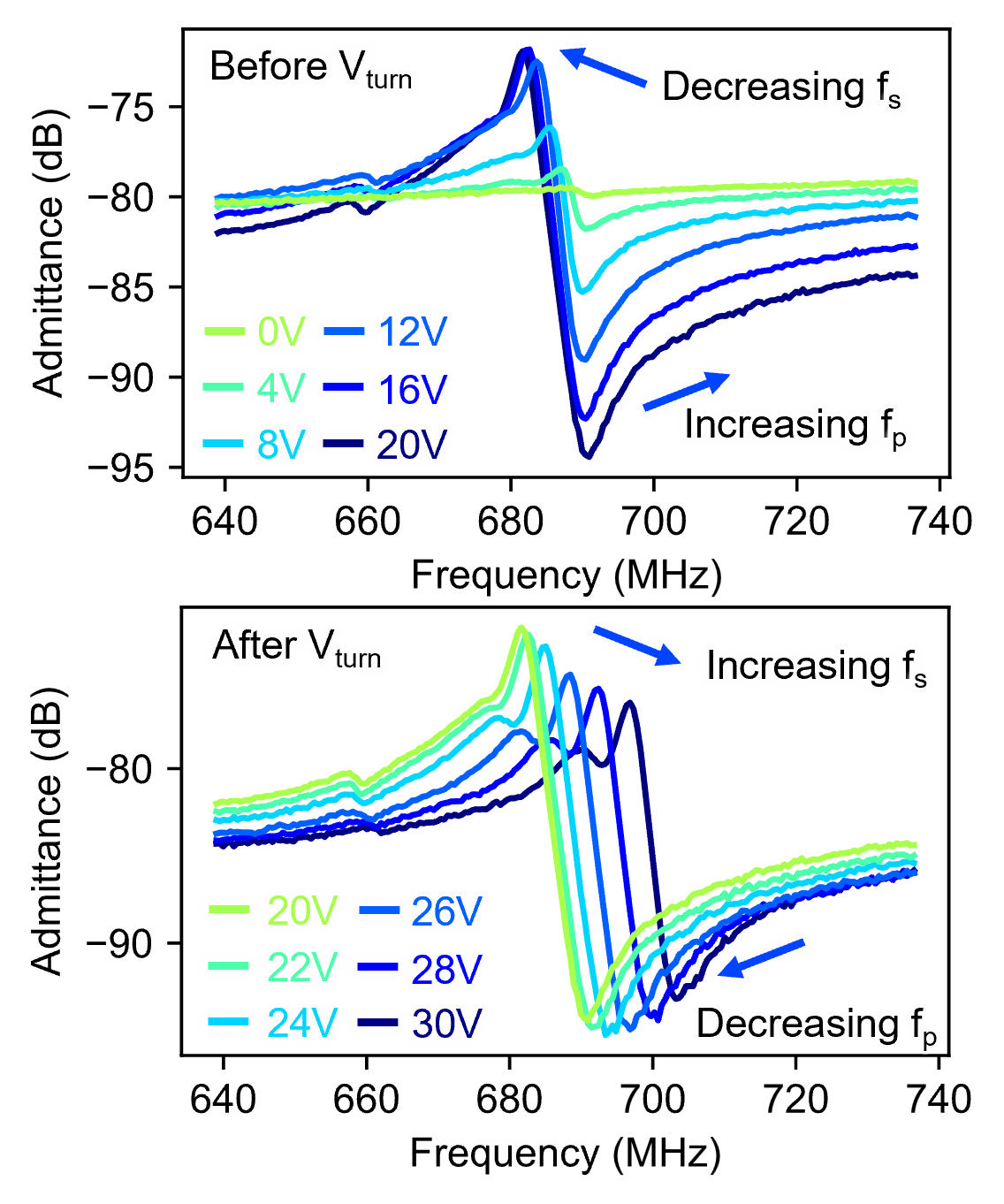}
\caption{(a) Admittance plots before $V_{turn}$ showing increasing figures of merit, and (b) after $V_{turn}$ showing reverse behavior.}
\label{fig1}
\end{figure}

To demonstrate device tunability, Fig. 5 shows device performance metrics for the two modes as a function of the applied DC bias. In both modes at 300 MHz and 700 MHz, similar trends are observed. As you apply larger and larger DC bias, the series resonance frequency is tuned through the change in electromechanical coupling, which only shifts the series resonance frequency \cite{defay_tunability_2011}. For this reason, the resonance frequency drops in a seemingly linear fashion. Along with the drop in resonance frequency, we observe an increase in the coupling between the modes and in the series quality factor ($Q$) (where we focus on series values, as they have lower impedance). Quality factors are determined using 3-dB bandwidths, and coupling is determined using Eq. 1 as:

\begin{equation}
    k^2 = \frac{\pi^2}{8}\cdot ( \frac{f_p^2}{f_s^2}-1)
\label{eq:Coupling}
\end{equation}

These trends continue until approximately 20 V, at which point we observe a dramatic change in behavior in the opposite direction for each resonance and value. The series resonance frequency turns up at a much larger rate, and the electromechanical coupling seems to drop significantly. Because this DC Bias is necessary for device excitation, increasing the number of electrodes will cancel the stress profile and yield zero coupling; therefore, only a two-electrode configuration is possible for the current electrode duty cycle.

Fig. 6 shows admittance plots for the 700 MHz mode before and after the voltage turning point. It is evident that the modes become more prominent below 20 V as the voltage increases. Subsequently, after 20 V, modes appear to decrease in prominence, with much larger changes in performance for a given voltage change. It is evident that the parallel resonance frequency changes substantially after this turning voltage. Before this point, the series resonance changes much more than the parallel resonance due to the increasing $k^2$, while changes in effective stiffness only slightly alter resonance frequencies \cite{rosenbaum_bulk_1988}. Above 20 V, both series and parallel resonance frequencies change dramatically. It is thought that, given the sudden onset, this change is due to electrostriction, though it could also originate from general material or electrode breakdown, given the drop in coupling \cite{vendik_modeling_2008,lanz_piezoelectric_2004}. It can also be seen that the capacitance varies markedly with the applied voltage, with lower capacitance at higher voltages. This is consistent with the expected decrease in permittivity, which also alters the resonance characteristics.

\begin{table}[t]
\caption{State-of-the-art ferroelectric resonators}
\label{tab:soa}
\centering
\footnotesize
\setlength{\tabcolsep}{4pt}
\renewcommand{\arraystretch}{1.12}
\begin{tabular}{@{}c c c c c c c c@{}}
\toprule
Ref. & Material & Excitation & $f$ (GHz) & $Q$ & $k^{2}$ & Tuning &
\makecell[c]{On-chip\\multi-$f$} \\
\midrule
\cite{alzuaga_tunable_2014} & STO   & Lateral   & 2.1  & 2400 & 0.02\% & 0.7\% & Y \\
\cite{wang_film_2020}       & ScAlN & Thickness & 2.9  & 210  & 18.1\% & 1.1\% & N \\
\cite{koohi_high_2017}      & BST   & Thickness & 2.1  & 100  & 8.6\%  & N/A   & N \\
\cite{lee_intrinsically_2013} & BTO & Thickness & 1.65 & 160  & 2\%    & 1.8\% & N \\
\cite{sis_intrinsically_2016} & BST & Thickness & 0.75 & 99  & 0.1\%    & N/A & N \\
\textbf{Here} & \textbf{BTO} & \textbf{Lateral} & \textbf{0.7} & \textbf{150} & \textbf{3\%} & \textbf{1.1\%} & \textbf{Y} \\
\bottomrule
\end{tabular}
\end{table}

We compare our results with those from ferroelectric acoustic resonators reported in the literature. Compared with other ferroelectric resonators, we exhibit comparable frequency, coupling, and frequency tunability, which we define as the percentage change in frequency as in Ref. \cite{lee_intrinsically_2013}. Our devices are also the only ones that are both monolithic, and have lithographically defined frequency setting without requiring bottom electrodes. This enables the fabrication of multiple devices at different frequencies without altering the film thickness.

\section{Conclusion}
Here, we demonstrate high-quality monolithic BTO acoustic resonators via lateral excitation of Lamb modes in released devices. Devices exhibit good tunability and acoustic performance up to 20 V, with several overmodes corresponding to S0-mode resonances. Compared with prior BTO works, we demonstrate competitive $Q$, coupling, and tunability without bottom electrodes. Future work will focus on larger devices with lower impedance and thicker films for higher $Q$.

\section*{Acknowledgment}

This work was supported by the NASA Space Technology Graduate Research Opportunity (NSTGRO), National Science Foundation (NSF) under CAREER Award No.\ 2339731, and the Office of Naval Research (Grant No. N00014-24-1–2063).

\balance
\bibliographystyle{IEEEtran}
\bibliography{IEEESettings,references}

\begin{thebibliography}{10}
\providecommand{\url}[1]{#1}
\csname url@samestyle\endcsname
\providecommand{\newblock}{\relax}
\providecommand{\bibinfo}[2]{#2}
\providecommand{\BIBentrySTDinterwordspacing}{\spaceskip=0pt\relax}
\providecommand{\BIBentryALTinterwordstretchfactor}{4}
\providecommand{\BIBentryALTinterwordspacing}{\spaceskip=\fontdimen2\font plus
\BIBentryALTinterwordstretchfactor\fontdimen3\font minus \fontdimen4\font\relax}
\providecommand{\BIBforeignlanguage}[2]{{%
\expandafter\ifx\csname l@#1\endcsname\relax
\typeout{** WARNING: IEEEtran.bst: No hyphenation pattern has been}%
\typeout{** loaded for the language `#1'. Using the pattern for}%
\typeout{** the default language instead.}%
\else
\language=\csname l@#1\endcsname
\fi
#2}}
\providecommand{\BIBdecl}{\relax}
\BIBdecl

\bibitem{attar_5g_2022}
H.~Attar, H.~Issa, J.~Ababneh, M.~Abbasi, A.~A.~A. Solyman, M.~Khosravi, and R.~Said~Agieb, ``\BIBforeignlanguage{en}{{5G} {System} {Overview} for {Ongoing} {Smart} {Applications}: {Structure}, {Requirements}, and {Specifications}},'' \emph{\BIBforeignlanguage{en}{Computational Intelligence and Neuroscience}}, vol. 2022, pp. 1--11, Oct. 2022.

\bibitem{hagelauer_microwave_2023}
A.~Hagelauer, R.~Ruby, S.~Inoue, V.~Plessky, K.-Y. Hashimoto, R.~Nakagawa, J.~Verdu, P.~D. Paco, A.~Mortazawi, G.~Piazza, Z.~Schaffer, E.~T.-T. Yen, T.~Forster, and A.~Tag, ``From {Microwave} {Acoustic} {Filters} to {Millimeter}-{Wave} {Operation} and {New} {Applications},'' \emph{IEEE Journal of Microwaves}, vol.~3, no.~1, pp. 484--508, Jan. 2023.

\bibitem{jackson_optical_1985}
K.~Jackson, S.~Newton, B.~Moslehi, M.~Tur, C.~Cutler, J.~Goodman, and H.~Shaw, ``\BIBforeignlanguage{en}{Optical {Fiber} {Delay}-{Line} {Signal} {Processing}},'' \emph{\BIBforeignlanguage{en}{IEEE Transactions on Microwave Theory and Techniques}}, vol.~33, no.~3, pp. 193--210, Mar. 1985.

\bibitem{qian_heterogeneous_2025}
F.~Qian, J.~Zheng, J.~Xu, and Y.~Yang, ``Heterogeneous {Interface}-{Enhanced} {Thin}-{Film} {SAW} {Devices} {Using} {Lithium} {Niobate} on {Si},'' \emph{IEEE Microwave and Wireless Technology Letters}, vol.~35, no.~1, pp. 123--126, Jan. 2025.

\bibitem{rinaldi_super-high-frequency_2010}
M.~Rinaldi, C.~Zuniga, {Chengjie Zuo}, and G.~Piazza, ``Super-high-frequency two-port {AlN} contour-mode resonators for {RF} applications,'' \emph{IEEE Transactions on Ultrasonics, Ferroelectrics and Frequency Control}, vol.~57, no.~1, pp. 38--45, Jan. 2010.

\bibitem{lu_rf_2021}
R.~Lu and S.~Gong, ``{RF} acoustic microsystems based on suspended lithium niobate thin films: advances and outlook,'' \emph{Journal of Micromechanics and Microengineering}, vol.~31, no.~11, p. 114001, Nov. 2021.

\bibitem{hummel_reconfigurable_2015}
G.~Hummel, Y.~Hui, and M.~Rinaldi, ``Reconfigurable {Piezoelectric} {MEMS} {Resonator} {Using} {Phase} {Change} {Material} {Programmable} {Vias},'' \emph{Journal of Microelectromechanical Systems}, vol.~24, no.~6, pp. 2145--2151, Dec. 2015.

\bibitem{devitt_spin-wave_2026}
C.~Devitt, S.~Tiwari, B.~Zivasatienraj, and S.~A. Bhave, ``\BIBforeignlanguage{en}{Spin-wave band-pass filters for {6G} communication},'' \emph{\BIBforeignlanguage{en}{Nature}}, Feb. 2026.

\bibitem{konno_tunable_2013}
A.~Konno, H.~Hirano, M.~Inaba, K.-y. Hashimoto, M.~Esashi, and S.~Tanaka, ``Tunable {Surface} {Acoustic} {Wave} {Filter} {Using} {Integrated} {Micro}-{Electro}-{Mechanical}-{System} {Based} {Varactors} {Made} of {Electroplated} {Gold},'' \emph{Japanese Journal of Applied Physics}, vol.~52, no.~7S, p. 07HD13, Jul. 2013.

\bibitem{nordquist_off_2013}
C.~D. Nordquist, R.~H. Olsson, S.~M. Scott, D.~W. Branch, T.~Pluym, and V.~Yarberry, ``On/{Off} micro-electromechanical switching of {AlN} piezoelectric resonators,'' in \emph{2013 {IEEE} {MTT}-{S} {International} {Microwave} {Symposium} {Digest} ({MTT})}.\hskip 1em plus 0.5em minus 0.4em\relax Seattle, WA, USA: IEEE, Jun. 2013, pp. 1--4.

\bibitem{alzuaga_tunable_2014}
S.~Alzuaga, W.~Daniau, R.~Salut, T.~Baron, S.~Ballandras, and E.~Defay, ``\BIBforeignlanguage{en}{Tunable and high quality factor {SrTiO}$_{\textrm{3}}$ surface acoustic wave resonator},'' \emph{\BIBforeignlanguage{en}{Applied Physics Letters}}, vol. 105, no.~6, p. 062901, Aug. 2014.

\bibitem{wang_film_2020}
J.~Wang, M.~Park, S.~Mertin, T.~Pensala, F.~Ayazi, and A.~Ansari, ``A {Film} {Bulk} {Acoustic} {Resonator} {Based} on {Ferroelectric} {Aluminum} {Scandium} {Nitride} {Films},'' \emph{Journal of Microelectromechanical Systems}, vol.~29, no.~5, pp. 741--747, Oct. 2020.

\bibitem{fichtner_alscn_2019}
S.~Fichtner, N.~Wolff, F.~Lofink, L.~Kienle, and B.~Wagner, ``\BIBforeignlanguage{en}{{AlScN}: {A} {III}-{V} semiconductor based ferroelectric},'' \emph{\BIBforeignlanguage{en}{Journal of Applied Physics}}, vol. 125, no.~11, p. 114103, Mar. 2019.

\bibitem{kim_crystal_2025}
H.~Kim, S.~Mathews, J.~Tischler, A.~B. Posadas, A.~A. Demkov, and A.~Piqué, ``\BIBforeignlanguage{en}{Crystal domain orientation control of epitaxial {BaTiO3} films integrated on silicon for large electro-optic response},'' \emph{\BIBforeignlanguage{en}{Applied Physics Letters}}, vol. 127, no.~5, p. 051102, Aug. 2025.

\bibitem{dong_monolithic_2023}
Z.~Dong, A.~Raju, A.~B. Posadas, M.~Reynaud, A.~A. Demkov, and D.~M. Wasserman, ``\BIBforeignlanguage{en}{Monolithic {Barium} {Titanate} {Modulators} on {Silicon}-on-{Insulator} {Substrates}},'' \emph{\BIBforeignlanguage{en}{ACS Photonics}}, vol.~10, no.~12, pp. 4367--4376, Dec. 2023.

\bibitem{kim_nature_2023}
I.~Kim, T.~Paoletta, and A.~A. Demkov, ``\BIBforeignlanguage{en}{Nature of electro-optic response in tetragonal {BaTiO}$_{\textrm{3}}$},'' \emph{\BIBforeignlanguage{en}{Physical Review B}}, vol. 108, no.~11, p. 115201, Sep. 2023.

\bibitem{koohi_high_2017}
M.~Z. Koohi, S.~Lee, and A.~Mortazawi, ``High {Q}$_{\textrm{m}}$×k$_{\textrm{t}}$$^{\textrm{2}}$ intrinsically switchable {BST} thin film bulk acoustic resonators,'' in \emph{2017 {IEEE} {MTT}-{S} {International} {Microwave} {Symposium} ({IMS})}.\hskip 1em plus 0.5em minus 0.4em\relax Honololu, HI, USA: IEEE, Jun. 2017, pp. 296--299.

\bibitem{lee_intrinsically_2013}
V.~Lee, S.~A. Sis, J.~D. Phillips, and A.~Mortazawi, ``Intrinsically {Switchable} {Ferroelectric} {Contour} {Mode} {Resonators},'' \emph{IEEE Transactions on Microwave Theory and Techniques}, vol.~61, no.~8, pp. 2806--2813, Aug. 2013.

\bibitem{kongbrailatpam_switchable_2025}
S.~S. Kongbrailatpam, R.~Akhil T~S, R.~James K~C, and G.~Pillai, ``\BIBforeignlanguage{en}{Switchable and tuneable high-performance acoustic modes in the {L}-{X} band using ferroelectric thin film on sapphire},'' \emph{\BIBforeignlanguage{en}{Microsystems \& Nanoengineering}}, vol.~11, no.~1, p. 217, Nov. 2025.

\bibitem{lee_nanoscale_2023}
D.~Lee, S.~Jahanbani, J.~Kramer, R.~Lu, and K.~Lai, ``\BIBforeignlanguage{en}{Nanoscale imaging of super-high-frequency microelectromechanical resonators with femtometer sensitivity},'' \emph{\BIBforeignlanguage{en}{Nature Communications}}, vol.~14, no.~1, p. 1188, Mar. 2023.

\bibitem{nordlander_ferroelectric_2020}
J.~Nordlander, F.~Eltes, M.~Reynaud, J.~Nürnberg, G.~De~Luca, D.~Caimi, A.~A. Demkov, S.~Abel, M.~Fiebig, J.~Fompeyrine, and M.~Trassin, ``\BIBforeignlanguage{en}{Ferroelectric domain architecture and poling of {BaTiO}$_{\textrm{3}}$ on {Si}},'' \emph{\BIBforeignlanguage{en}{Physical Review Materials}}, vol.~4, no.~3, p. 034406, Mar. 2020.

\bibitem{vasudevan_domain_2023}
A.~T. Vasudevan and S.~K. Selvaraja, ``\BIBforeignlanguage{en}{Domain effects on the electro-optic properties of thin-film barium titanate},'' \emph{\BIBforeignlanguage{en}{Optical Materials Express}}, vol.~13, no.~4, p. 956, Apr. 2023.

\bibitem{defay_tunability_2011}
E.~Defaÿ, N.~B. Hassine, P.~Emery, G.~Parat, J.~Abergel, and A.~Devos, ``Tunability of {Alluminum} {Nitride} {Acoustic} {Resonators}: {A} {Phenomenological} {Approach},'' \emph{IEEE Transactions on Ultrasonics, Ferroelectrics, and Frequency Control}, vol.~58, no.~12, pp. 2516--2520, Dec. 2011.

\bibitem{rosenbaum_bulk_1988}
J.~F. Rosenbaum, \emph{\BIBforeignlanguage{eng}{Bulk acoustic wave theory and devices}}, ser. The {Artech} {House} acoustics library.\hskip 1em plus 0.5em minus 0.4em\relax Boston: Artech House, 1988.

\bibitem{vendik_modeling_2008}
I.~B. Vendik, P.~A. Turalchuk, O.~G. Vendik, and J.~Berge, ``\BIBforeignlanguage{en}{Modeling tunable bulk acoustic resonators based on induced piezoelectric effect in {BaTiO}$_{\textrm{3}}$ and {Ba}$_{\textrm{0.25}}${Sr}$_{\textrm{0.75}}${TiO}$_{\textrm{3}}$ films},'' \emph{\BIBforeignlanguage{en}{Journal of Applied Physics}}, vol. 103, no.~1, p. 014107, Jan. 2008.

\bibitem{lanz_piezoelectric_2004}
R.~Lanz, ``\BIBforeignlanguage{en}{Piezoelectric thin films for bulk acoustic wave resonator applications : from processing to microwave filters},'' Ph.D. dissertation, Lausanne, EPFL, 2004.

\bibitem{sis_intrinsically_2016}
S.~A. Sis, S.~Lee, V.~Lee, and A.~Mortazawi, ``An {Intrinsically} {Switchable}, {Monolithic} {BAW} {Filter} {Using} {Ferroelectric} {BST},'' \emph{IEEE Microwave and Wireless Components Letters}, vol.~26, no.~1, pp. 25--27, Jan. 2016.

\end{thebibliography}

\end{document}